\documentclass[twocolumn]{aastex62}

\usepackage{graphicx}
\usepackage{subfigure}
\usepackage{amsmath}
\usepackage{epsfig}

\newcommand{\refsec}[1]{Section~\ref{#1}}
\newcommand{\reffig}[1]{Fig.~\ref{#1}}

\newcommand{\gasoline}{\texttt{GASOLINE}}
\newcommand{\kms}{km s$^{-1}$}
\newcommand{\Illustris}{\texttt{Illustris}}

\shorttitle{cEs formation}
\shortauthors{Du et al.}

\begin{document}
\title{The formation of compact elliptical galaxies in the vicinity of a massive galaxy: the role of ram-pressure confinement}

\correspondingauthor{Min Du and Victor P. Debattista}
\email{dumin@pku.edu.cn; vpdebattista@uclan.ac.uk}

\author{Min Du}
\affil{Kavli Institute for Astronomy and Astrophysics, Peking University, Beijing 100871, China}

\author{Victor P. Debattista}
\affiliation{Jeremiah Horrocks Institute, University of Central Lancashire, Preston PR1 2HE, UK}

\author{Luis C. Ho}
\affiliation{Kavli Institute for Astronomy and Astrophysics, Peking University, Beijing 100871, China}
\affiliation{Department of Astronomy, School of Physics, Peking University, Beijing 100871, China}

\author{Patrick C$\hat{o}$t$\acute{e}$}
\affiliation{Herzberg Institute of Astrophysics, National Research Council of Canada, Victoria, BC V9E 2E7, Canada}

\author{Chelsea Spengler}
\affiliation{Department of Physics and Astronomy, University of Victoria, Victoria, BC V8P 5C2, Canada}

\author{Peter Erwin}
\affiliation{Max-Planck-Insitut f$\ddot{u}$r extraterrestrische Physik, Giessenbachstrasse, 85748 Garching, Germany}

\author{James W. Wadsley}
\affiliation{Department of Physics \& Astronomy, McMaster University, Hamilton, L8S 4M1, Canada}

\author{Mark A. Norris}
\affiliation{Jeremiah Horrocks Institute, University of Central Lancashire, Preston PR1 2HE, UK}

\author{Samuel W. F. Earp}
\affiliation{Jeremiah Horrocks Institute, University of Central Lancashire, Preston PR1 2HE, UK}

\author{Thomas R. Quinn}
\affiliation{Astronomy Department, University of Washington, Seattle, Washington, USA}

\author{Karl Fiteni}
\affiliation{Department of Physics, University of Malta, Msida MSD 2080, Malta}
\affiliation{Institute of Space Sciences \& Astronomy, University of Malta, Msida MSD 2080, Malta}

\author{Joseph Caruana}
\affiliation{Department of Physics, University of Malta, Msida MSD 2080, Malta}
\affiliation{Institute of Space Sciences \& Astronomy, University of Malta, Msida MSD 2080, Malta}

\begin{abstract}
Compact ellipticals (cEs) are outliers from the scaling relations of
early-type galaxies, particularly the mass-metallicity relation which
is an important outcome of feedback. The formation of such low-mass,
but metal-rich and compact, objects is a long-standing puzzle. Using a
pair of high-resolution $N$-body+gas simulations, we investigate the
evolution of a gas-rich low-mass galaxy on a highly radial orbit
around a massive host galaxy. As the infalling low-mass galaxy passes
through the host's corona at supersonic speeds, its diffuse gas
outskirts are stripped by ram pressure, as expected. However, the
compactness increases rapidly because of bursty star formation in the
gas tidally driven to the centre. The metal-rich gas produced by
supernovae and stellar winds is confined by the ram pressure from the
surrounding environment, leading to subsequent generations of stars
being more metal-rich. After the gas is depleted, tidal interactions
enhance the metallicity further via the stripping of weakly bound, old
and metal-poor stars, while the size of the satellite is changed only
modestly. The outcome is a metal-rich cE that is consistent with
observations. These results argue that classical cEs are neither the stripped
remnants of much more massive galaxies nor the merger remnants of normal dwarfs. 
We present observable predictions that can be used to test
our model.
\end{abstract}


\section{Introduction}
\label{sec:intro}

At the low-mass end of the early-type galaxy population, the
well-known mass/luminosity-size relation \citep{Larson1981} splits
into diffuse and compact branches. The compact branch is composed of
compact ellipticals (cEs) and may even extend to the regime of
ultra-compact dwarfs \citep{Mieske2005, Hasegan2005, Chilingarian2007,
Chilingarian2009, Castelli2008, Price2009, Huxor2011, Huxor2013,
Norris2014, Chilingarian2015, Guerou2015}. cEs have effective
(half-mass) radii ($R_e$) that are generally less than 0.6 kpc, while their
diffuse counterparts, the dwarf elliptical galaxies (dEs) or dwarf
spheroidals (dSphs), have $R_e\sim 0.6-3$ kpc at similar masses. One cE
formation scenario proposes that cEs are low-mass classical
ellipticals, which is supported by the fact that they follow the same
trend on the fundamental plane as the giant ellipticals
\citep[e.g.][]{Wirth1984, Kormendy2009, Kormendy2012,
Paudel2014a}. This implies formation through hierarchical mergers, as
in ``normal'' ellipticals. Most cEs are notably more metal-rich than
dEs \citep[e.g.][]{Chilingarian2009, Francis2012, Janz2016} and are
outliers from the mass-metallicity relation of massive early-type
galaxies \citep[e.g.][]{Gallazzi2005, Panter2008, Thomas2010} and
low-mass galaxies in the Local Group
\citep{Kirby2013}. Their dramatic offset from the mass-metallicity
relation is an extraordinary challenge for the merger scenario.

An alternative formation scenario addresses the problem of high
metallicity by proposing that cEs are the remnants of larger, more
massive galaxies \citep[e.g.][]{Faber1973, Bekki2001b, Choi2002,
Graham2002}. In this scenario, their disks are stripped via strong
tidal interactions with an even more massive host galaxy, leaving only
the compact, metal-rich bulges. For instance, the nearest cE, and
prototype of the class, M32, is known to be interacting with the
Andromeda galaxy (M31); \citet{Choi2002} showed that M32's outer
isophotes are distorted by M31's tidal field. However, the offset from
the mass-metallicity relation of the dIrrs and dSphs is $\sim 0.4-0.6$
dex, \citep{Gallazzi2005, Kirby2013}, requiring a Milky Way-mass
progenitor, whereas the total stellar halo of M31 is much less massive
\citep[$\sim 10^9M_\odot$ level; e.g.][]{Ibata2007, Courteau2011,
Gilbert2012, Ibata2014}. Thus, no evidence for such a massive
progenitor to M32 exists in M31. It is also worth noting that only the
innermost regions (within $R_e/8$) of massive, early-type galaxies
show metallicities comparable to those of cEs \citep{McDermid2015}.
Furthermore, by simulating the interaction between M31 and M32, 
\citet{Dierickx2014} showed that stripping from
plausible progenitors was probably not sufficiently efficient. This is 
consistent with the kinematic measurement of M32 \citep{Howley2013}, 
which showed no evidence of tidal stripping within 1 kpc.

A dense cluster environment has usually been believed to play an
essential role in generating cEs, because many of them are found in
clusters. Recently, however, some cEs have been found in groups
\citep{Norris2014, Janz2016, Ferre-Mateu2018} and even in the field
\citep{Huxor2013, Paudel2014a, Chilingarian2015}. The systematic study
of compact low-mass galaxies in \citet{Norris2014} showed that cEs
exist in a wide variety of environments. They suggested that the
formation of cEs is likely to be associated with an adjacent massive
host, with cEs in the field having been flung out of bound systems via
three-body interactions \citep[e.g.][]{Chilingarian2015}.

In this paper, we report on a numerical simulation specifically
designed to explore the origin of cE galaxies around a large host via
highly elliptical orbits. We simulate the evolution of a low-mass
galaxy on a highly radial orbit around a massive disk galaxy. As a
low-mass satellite falls into the hot corona of its host, the supply
of metal-poor fresh gas is cut off, a process referred to as
strangulation \citep{Peng2010, Peng2015}. A metal-poor dwarf might be
expected to result. Here, we show that instead a low-mass satellite on
such an orbit evolves into a metal-rich cE after several close flybys
through the dense corona of its host.

The paper is organized as follows. \refsec{sec:sim} describes the
simulation setup. In \refsec{sec:result}, we show the evolution of the
morphology and kinematics of the low-mass satellite. Its rapid
chemical enrichment due to ram-pressure confinement is presented in
\refsec{sec:gas}. In \refsec{sec:dis}, we discuss the application of 
our work, the difference from previous simulations, and observational 
predictions. Our conclusions and summary of our results are 
presented in \refsec{sec:sum}.


\section{The simulation}
\label{sec:sim}

\begin{figure*}[htbp]
\begin{center}
\subfigure{\includegraphics[width=0.8\textwidth]{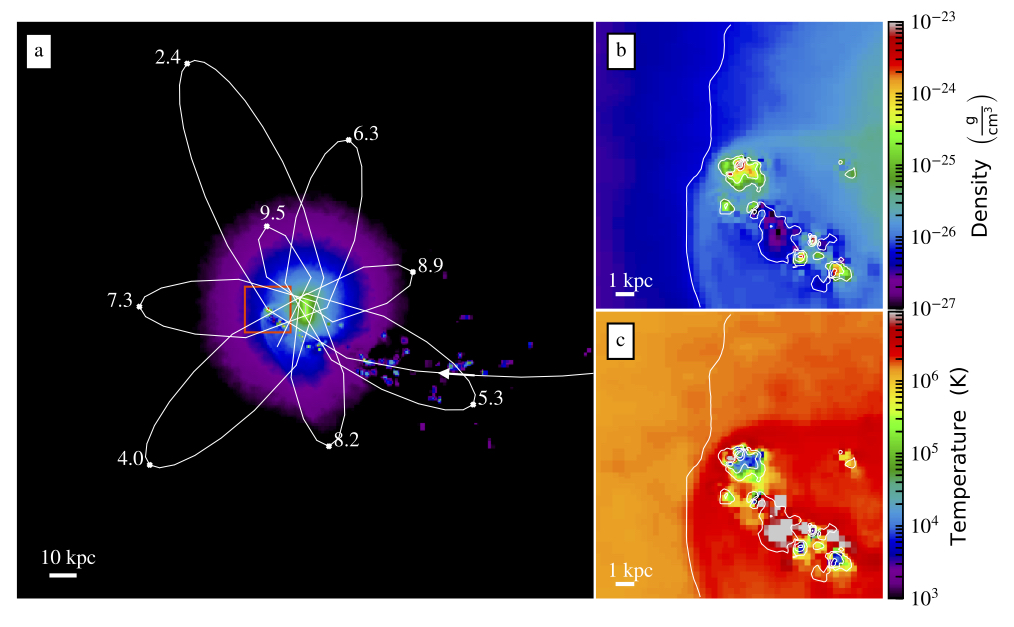}}
\caption{Panel (a) shows the trajectory of the low-mass satellite galaxy 
overlaid on a narrow slice of gas density at $t=1.66$ Gyr. The white
dots mark the apocenters. The white arrow represents the position and
motion of the satellite at $t=1.37$ Gyr. At this point, both the host
galaxy and the satellite galaxy are viewed nearly edge-on. While still
more than $100$~kpc from the centre of the host, the satellite has a
stellar mass of $2.6\times 10^8 M_\odot$, and has attained a velocity
with respect to the host of $\thickapprox 210$ \kms. Panels (b) and
(c) are zoomed-in plots of the infalling satellite marked by the red
square region in panel (a) at $t=1.66$ Gyr. Panels (b) and (c) show
the density and temperature of gas, respectively, overlaid with gas
density contours. Panel (b) uses the same color bar as panel (a). At
$t\sim 1.6$ Gyr, the satellite first passes the pericenter at $\sim 16$
kpc from the center of the host. A bow shock is clearly visible in
front of the satellite due to its supersonic bulk motion relative to
the host corona. It never merges into the massive galaxy during the
simulation, which lasts for 10 Gyr.}
\label{fig:traj}
\end{center}
\end{figure*} 

\begin{figure}[htbp]
\begin{center}
\subfigure{\includegraphics[width=0.48\textwidth]{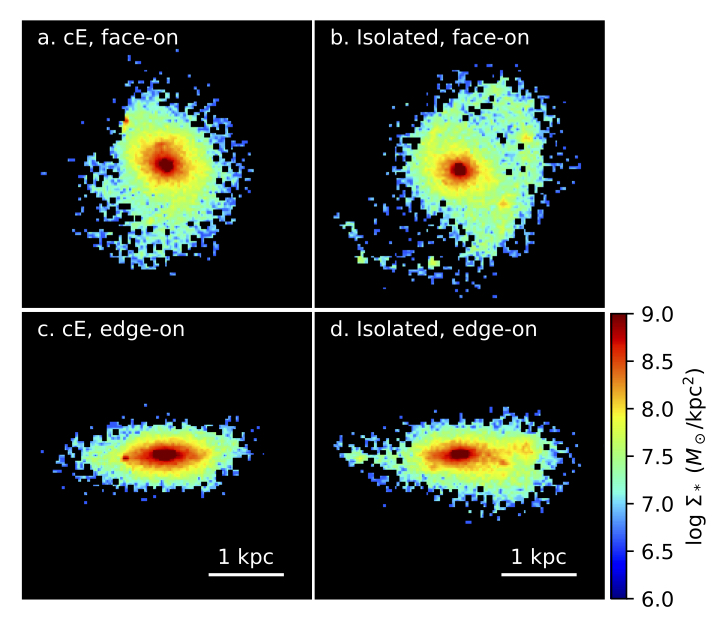}}
\caption{Stellar surface density distributions of the cE (left) and 
isolated (right) models at $t=0.9$ Gyr in the face-on (top) and
edge-on (bottom) views. At this time, the cE has $M_* \simeq 2.0\times
10^8 M_\odot$ and $R_{e, *}\simeq 307$ pc, while the isolated galaxy
has $M_* \simeq 1.8 \times 10^8 M_\odot$, $R_{e, *} \simeq 446$ kpc. }
\label{fig:SurfDens100}
\end{center}
\end{figure}

\begin{figure}[htbp]
\begin{center}
\subfigure{\includegraphics[width=0.48\textwidth]{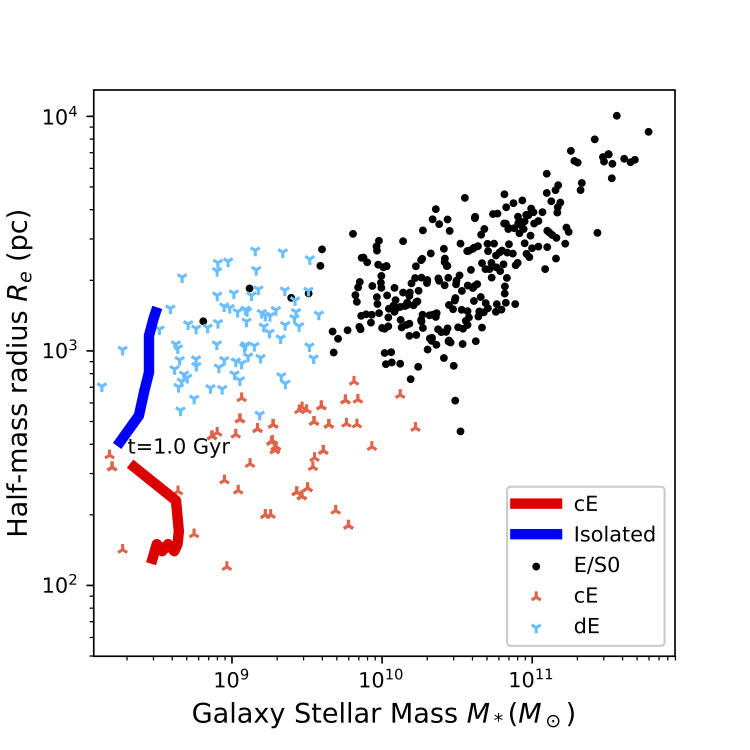}}
\caption{Evolution of the mass and size of the cE and isolated models 
overlaid on the observed mass-size relation. The data of elliptical/lenticular galaxies (E/S0s, black
dots), cEs (red stars), and dEs (blue stars) are adopted from
\citet{Janz2016}. At $t=1$ Gyr, the dwarf satellite has a mass
and size similar to the isolated model. Its evolution in the vicinity of the
massive host transforms it into a compact object.}
\label{fig:MassSize}
\end{center}
\end{figure}

\begin{figure}[htbp]
\begin{center}
\subfigure{\includegraphics[width=0.48\textwidth]{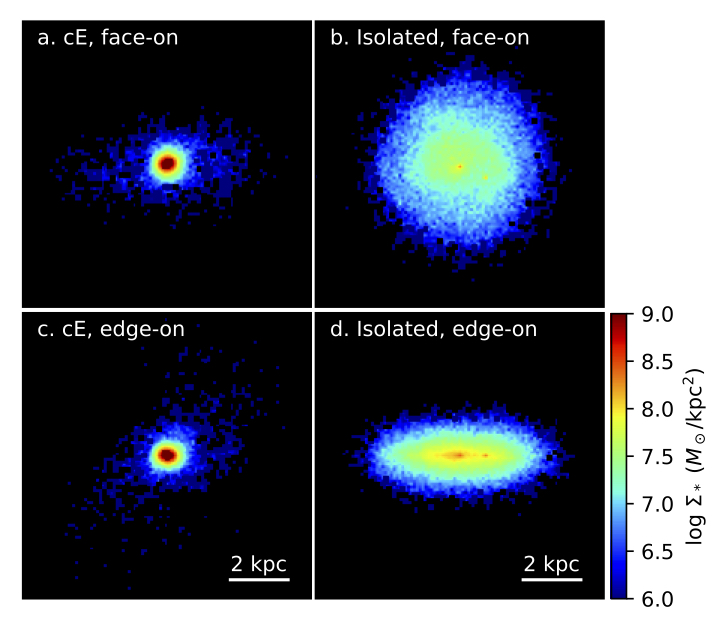}}
\caption{Stellar surface density distributions of the cE (left) and 
isolated (right) models at $t=10$ Gyr in the face-on (top) and edge-on
(bottom) views. The satellite galaxy evolves into a cE after several
close passages of the massive galaxy, while it becomes a normal
diffuse galaxy in isolation. At this time, the cE has $M_* \simeq
2.7\times 10^8 M_\odot$ and $R_{e, *}\simeq 144$ pc, while the
isolated galaxy has $M_* \simeq 3.1 \times 10^8 M_\odot$ and $R_{e, *}
\simeq 1.5$ kpc. Note the difference in their surface densities.}
\label{fig:SurfDens}
\end{center}
\end{figure}

We perform a simulation of a low-mass galaxy orbiting a massive
galaxy. The low-mass galaxy starts at 200 kpc from the larger galaxy
with its velocity set for it to fall to within $\sim 20$ kpc of the
host, giving it a highly eccentric orbit which brings it inside the
densest part of the hot gas corona. The motivation for a highly
eccentric orbit is two-fold. Foremost, cEs are relatively rare
objects, and therefore require unusual properties of some sort.
Second, M32 is very close to M31, which requires its orbit
to bring it close to the host.

Because we want to track the star formation and chemical enrichment of
the galaxies, both galaxies start out as dark matter halos with gas
but no star particles. The dark matter halos have Navarro-Frenk-White
(NFW) \citep{Navarro1996} profiles, truncated exponentially beyond the
virial radius. The dark matter halo is isotropic and is built using
Eddington inversion \citep{Kazantzidis2004}. The dark matter halo of
the massive host system has a virial radius of $r_\mathrm{vir} = 198$ kpc,
concentration of $c_\mathrm{vir} = 19$, and a virial mass of $M_\mathrm{vir} = 9.0 \times
10^{11} M_\odot$, while the low-mass system has $r_\mathrm{vir} = 54$
kpc, $c_\mathrm{vir} = 15$, and $M_\mathrm{vir} = 1.8 \times 10^{10} M_\odot$. The
halos of the host and low-mass galaxies are comprised of $5 \times
10^6$ and $10^5$ particles, respectively. The gas of the host galaxy
has a similar profile with just $10\%$ of the mass and is in thermal
equilibrium. Because we do not wish to set the morphology of the dwarf
by hand, we also include a corona for the dwarf galaxy; at this low
mass, its corona cools rapidly and largely settles into a disk. The
gas coronae have angular momenta such that the spin parameter $\lambda
\thickapprox 0.041$ for the massive host and $\lambda \thickapprox
0.02$ for the low-mass system. Initially, there is an equal number of
gas particles as dark particles. We use a spline softening of
$\epsilon = 103$ pc for dark matter particles and $\epsilon = 50$ pc
for gas and stars. The initial masses of gaseous and stellar particles 
are $1.8\times 10^5M_\odot$ and $9.3\times 10^3 M_\odot$, respectively.

We do not introduce any stars into the initial conditions. Rather, all
stars form self-consistently out of gas, ensuring that the chemical
evolution is not imposed by the initial conditions. For the main
galaxy these initial conditions correspond to the state of the galaxy
after the last major merger at redshift $z \sim 2$. Instead for the
dwarf galaxy, the corona quickly cools and settles into a disk. 
When the gas number density reaches $100$
cm$^{-3}$ and the temperature falls below $15,000$ K in a converging
flow, stars can form with a probability of $10\%$ per dynamical
time. A gas particle that loses more than $79\%$ of its initial
mass to star formation will disperse its mass amongst its nearest
neighbours and be removed. Each star particle represents an entire stellar
population with a Miller-Scalo initial mass function
\citep{Miller1979}. Supernovae and stellar winds inject energy and
metallicity into the interstellar medium (ISM) using the blastwave
prescription of \citet{Stinson2006}. Following \citet{Governato2010},
$0.4 \times 10^{51}$ erg is returned from each supernova. The gas
starts with zero metallicity. Then, iron and $\alpha$-elements are
produced in a subgrid model of Type Ia and II supernovae, and
asymptotic giant branch stars \citep{Stinson2006} according to the
yields of \citet{Raiteri1996}, \citet{Thielemann1986}, and
\citet{Weidemann1987}. We include metal and thermal diffusion, based
on a subgrid model of turbulence using the local smoothing length and
velocity gradients \citep{Smagorinsky1963, Wadsley2008}, with the
diffusion parameter set to 0.03.

The simulation is evolved for 10 Gyr using \gasoline2, a Smoothed
Particle Hydrodynamics (SPH) code \citep{Wadsley2004, Wadsley2017}. 
This improved SPH code avoids the insufficient mixing of gas 
caused by contact discontinuities \citep[see][]{Wadsley2017}.
We use a base time step of $\Delta t = 10$ Myr with time steps refined
such that $\delta t = \Delta t/2^n < \eta\sqrt{\epsilon/a_g}$, where
$a_g$ is the acceleration at a particle's position and the refinement
parameter $\eta = 0.175$. We set the opening angle of the tree-code
gravity calculation to $\theta = 0.7$. The time step of gas particles
also satisfies the condition $\delta t_\mathrm{gas} =
\eta_\mathrm{courant} h/[(1+\alpha)c + \beta \mu_\mathrm{max}]$, where
$\eta_\mathrm{courant} = 0.4$, $h$ is the SPH smoothing length set
over the nearest 32 particles, $\alpha = 1$ and $\beta = 2$ are, respectively,
the coefficients for the linear and quadratic terms of the artificial
viscosity, and $\mu_\mathrm{max}$ is described in \citet{Wadsley2004}.
In order to help determine the effect of environment on the dwarf, we
also evolve it in isolation.


\section{Evolution of the satellite galaxy}
\label{sec:result}

\reffig{fig:traj} (a) shows the trajectory of the satellite relative to 
the host galaxy, which forms an M31-like galaxy
\citep[e.g.][]{Debattista2017}; the evolution of the host will not be
discussed here. The orbit decays rapidly due to dynamical friction,
with its apocenter decreasing from $200$ to $\sim 120$ kpc after
the first pericentric passage and to $\sim 50$ kpc by the end of the
simulation.

The low-mass system evolved in isolation loses its gaseous corona at a
rate of $\sim 5\times 10^8 M_\odot/{\rm Gyr}$ ($\sim 25\%$ of the
total gas mass per Gyr) in the first 2 Gyr, due to cooling, star
formation and feedback. The loss of diffuse gas is even faster when
the low-mass system is evolved in the orbit around the massive galaxy.

Before $t=1$ Gyr, the satellite is still further than $150$ kpc from
the center of the massive host, because it moves slowly (tens of \kms\
level). Thus, the satellite evolves essentially by
itself. Fig. \ref{fig:SurfDens100} reveals that, as a result, the
satellite galaxy is very similar to the model evolved in
isolation. The satellite's gaseous corona quickly cools and star
formation commences. Within a sphere of 5 kpc, about $\sim 60\%$ of
the baryonic mass is cold gas (temperature $T<15, 000 K$) by $t=0.9$ Gyr
and a disky dwarf of stellar mass $\sim 2 \times 10^8 M_\odot$ forms
(see \reffig{fig:SurfDens100}).

The satellite is affected as it approaches the massive galaxy via a
number of mechanisms: (1) ram-pressure stripping of weakly bound gas
\citep{Gunn1972}, (2) tidal stripping of stars, (3) triggering of star
formation (SF) within the remaining gas, (4) stellar heating via tidal
stirring \citep{Mayer2001}.
As the satellite sinks deeper into the host corona, it experiences an
increased ram pressure and a very large fraction of gas in the diffuse
outskirts is rapidly stripped by the ram pressure
\citep[e.g.][]{Gunn1972}. The satellite reaches the first pericenter at
$\sim 1.6$ Gyr travelling at $\sim 400$ \kms; the sound speed in the
host's corona is $\sim 160$ \kms, meaning the satellite's motion is
highly supersonic. In panels (b) and (c) of \reffig{fig:traj}, we show
magnified density and temperature images around the dwarf. A bow-shock
structure \citep[e.g.][]{Domainko2006, Jachym2007} is distinguishable
in front of the satellite due to its supersonic bulk motion relative
to the corona. Nevertheless, the centre retains a reservoir of cold
gas.

This process repeats several times, with the orbit slowly decaying.
By $t\approx 6.5$ Gyr, after four pericentric passages, the satellite
galaxy has lost all of its gas and is completely quenched. The satellite
has a size typical of cEs for its mass, as shown in
\reffig{fig:MassSize}, while the isolated model has a size similar
to dEs. At $t=10$ Gyr, the satellite has reached a stellar mass $M_*
\simeq 2.7\times 10^8 M_\odot$ and a face-on effective radius of $R_{e,*} =
144$ pc. In comparison, the isolated model forms a normal low-mass
galaxy of $M_* \simeq 3.1 \times 10^8 M_\odot$ and $R_{e, *} \simeq
1.5$ kpc. The surface density distributions of the two models at 10
Gyr are compared in \reffig{fig:SurfDens}. The isolated model still
retains a cold gas reservoir of a mass of $3.5 \times 10^8 M_\odot$ within
$r<5$ kpc. Thus, the isolated model is a typical diffuse and gas-rich
dwarf galaxy.

\begin{figure}[htbp]
\begin{center}
\subfigure{\includegraphics[width=0.48\textwidth]{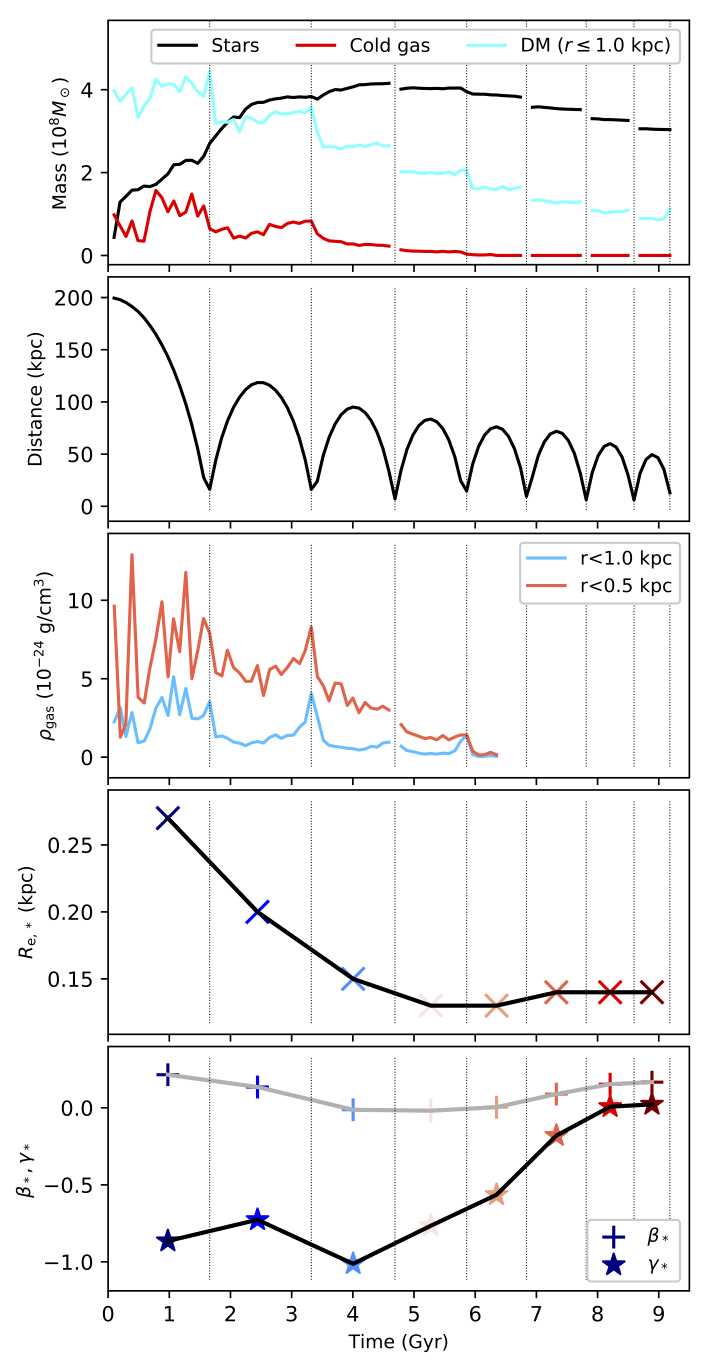}}
\caption{Evolution of the satellite galaxy. In the top panel, the black, 
red, and cyan lines represent the evolution of the total mass of
stars, cold gas ($T<15, 000$ K), and dark matter within
$r<1.0$ kpc, respectively. The second panel shows the separation
between the satellite and the host galaxy. The vertical dotted lines
mark the pericenters. The third panel shows the evolution of the cold
gas density, $\rho_\mathrm{gas}$. The red and blue tracks represent the
results obtained within a sphere of radius of 0.5 and 1.0 kpc,
respectively. The data during passages at $<10$ kpc are excluded as
they cannot be distinguished from the host galaxy. The fourth panel
shows the stellar effective radius ($R_{e, *}$) and the bottom panel
the velocity dispersion anisotropies of the satellite measured face-on
at apocenters. The blue and red markers in panels 4 and 5 represent
stages (1) and (2), respectively. For comparison, we include the data
point at $t=1.0$ Gyr when the satellite has not yet passed
pericentre.}
\label{fig:evo}
\end{center}
\end{figure}

\begin{figure}[htbp]
\begin{center}
\subfigure{\includegraphics[width=0.48\textwidth]{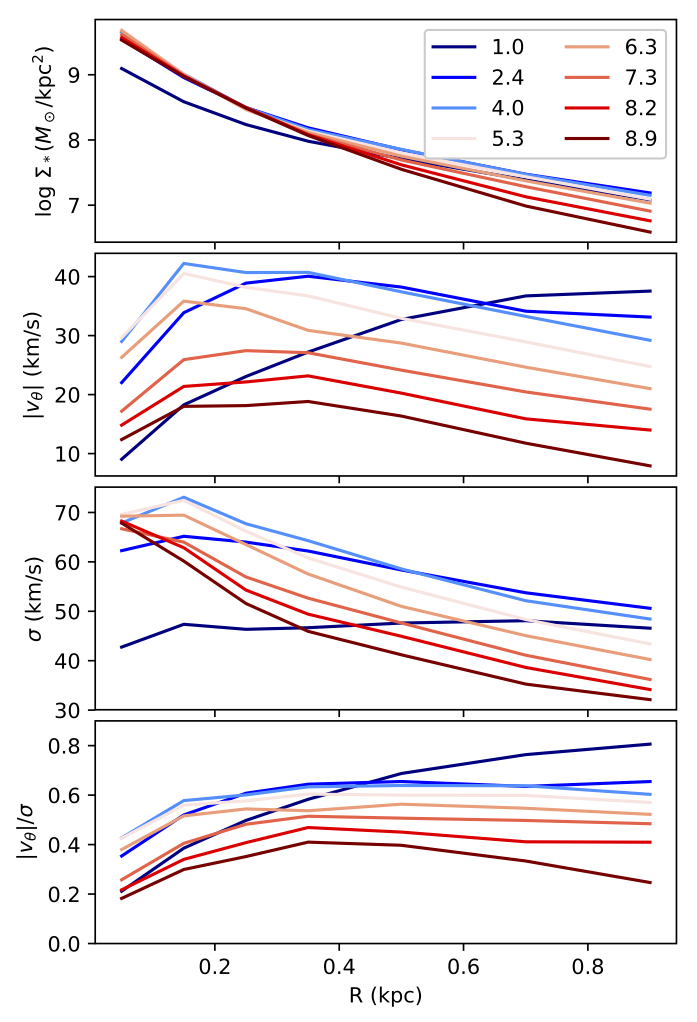}}
\caption{Radial profiles in cylindrical coordinates of the stellar 
surface density and kinematic properties of the satellite galaxy,
measured at apocenters. From top to bottom, we show the surface
density; mean rotation speed, $\vert v_\theta \vert$; velocity
dispersion, $\sigma=\sqrt{\sigma_R^2 + \sigma_\theta^2 + \sigma_z^2}$;
and the ratio $\vert v_\theta \vert/\sigma$. The blue and red series
of lines represent stages (1) and (2), respectively. The data point at
$t=1.0$ Gyr shows the properties before the first pericentric
passage.}
\label{fig:rot}
\end{center}
\end{figure}

\subsection{Morphology and kinematics}
\label{sec:MK}

\reffig{fig:evo} shows the evolution of some basic parameters of the 
satellite galaxy. The first panel presents the masses of stars, cold
gas ($T<1.5\times 10^4$ K), and dark matter. The second panel shows
the distance between the satellite and the massive host. The vertical
dotted lines in all of the panels mark the times of pericentric
passages. The third panel of \reffig{fig:evo} shows the mean density
of cold gas, $\rho_\mathrm{gas}$, within spheres of radii 0.5 and 1.0
kpc. The cold gas is compressed as the satellite moves close to its
massive host, triggering starbursts. The fourth panel shows the
effective radius, $R_{e, *}$. The fifth (bottom) panel shows the
velocity dispersion anisotropies, $\beta_*$ and $\gamma_*$, of the
satellite measured at the apocenter. We define the anisotropies of the
velocity dispersion ellipsoid as $\beta_* = 1-\sigma_\theta^2 /
\sigma_R^2$ and $\gamma_*=1-\sigma_z^2 /
\sigma_R^2$, where both $\beta_*$ and $\gamma_*$ approach 0 as the
system becomes isotropic. In \reffig{fig:rot}, we show the radial
kinematics at the apocenters. Figs. \ref{fig:evo} and \ref{fig:rot}
reveal two distinct stages in the evolution of the satellite: (1) the
compactness increases rapidly during 1 to 4 Gyr; (2) the mass and
rotation decrease gradually during $4-10$ Gyr. Stages (1) and (2)
correspond to the blue and red series of lines, respectively, in
\reffig{fig:rot}.
 
At $t=1$ Gyr, the diffuse stellar distribution leads to a low
$\sigma$ and a gently rising $\vert v_\theta \vert$
(\reffig{fig:rot}) with the radius. During stage (1), the compactness
increases rapidly as a compact core is built up (the top panel of
\reffig{fig:rot}) via rapid star formation at the centre. As 
suggested by previous simulations, both the high ram pressure
\citep{Bekki2003b, Kronberger2008, Nehlig2016, Henderson2016} and
strong tidal interactions \citep[e.g.][]{Smith2010, Renaud2014,
Nehlig2016} are likely to enhance the star formation rate (SFR; see
also \refsec{sec:BSF}) by compressing the gas in the central region
(see the third panel of \reffig{fig:evo}). At $t=4.5$ Gyr, $M_*$
reaches its maximum value, $4.5\times 10^8M_\odot$, while $R_{e, *}$
has dropped to $\thickapprox 160$ pc. Such a compact concentration of mass
results in a steeply rising rotation curve. The low-mass galaxy still
retains a significant rotation at $t=4.0$ Gyr, with $\vert v_\theta
\vert/\sigma \sim 0.6$. Therefore, {\it during stage (1), the infalling
low-mass galaxy is transformed into a gas-poor and compact, but
rapidly rotating, stellar system.} Strong tidal interactions have only
happened twice (at $t=1.7$ Gyr, and at $3.3$ Gyr) by this stage, but
the stellar mass lost due to tidal stripping is not the most important
factor driving the evolution of the dwarf up to this point, although
dark matter is substantially stripped.

The growth of the stellar mass gradually declines because the cold gas
is depleted, transitioning to stage (2). During this stage, $M_*$
decreases due to the loss of weakly bound stars via tidal
stripping. However, tidal stripping has only a modest effect on the
compactness, as shown by the small change in $R_{e, *}$ during 5-9 Gyr
(the fourth panel of \reffig{fig:evo}). The remnant at $t=10$ Gyr has
a similar mass, $M_* \simeq 2.7 \times 10^8 M_\odot$, as the
progenitor before its first pericentric passage. About $40\%$ of
the stellar mass is stripped with respect to the maximum $M_*$ at
$t=4.5$ Gyr. Thus, tidal stripping of stars certainly plays a role, but
it is not the dominant factor in the formation of this cE. The
satellite loses $\sim 75\%$ of its dark matter due to the tidal
interactions (the cyan profile in the top panel of \reffig{fig:evo}).

The kinematic properties and morphology of the infalling low-mass
galaxy are also significantly changed during this stage. Strong tidal
interactions heat stars impulsively via tidal stirring
\citep[e.g.][]{Mayer2001}, leading to a gradual decline in the
rotation (\reffig{fig:rot}) and the anisotropies (the bottom panel of
\reffig{fig:evo}). The satellite becomes a compact and kinematically isotropic 
stellar system dominated by random motions, with $\vert v_\theta \vert/\sigma
\sim 0.3$, by $t=8.9$ Gyr. We notice that M32 also has nearly isotropic 
orbital families, according to the result of the triaxial Schwarzschild modelling 
in \citet{vandenBosch2010}.

When the low-mass galaxy evolves in isolation, both its effective
radius and stellar mass increase continuously. It ends as a gas-rich,
diffuse, fast-rotating ($\vert v_\theta \vert/\sigma \sim 0.8$) galaxy
that has a stellar mass comparable to that of the infalling low-mass
galaxy.

\subsection{Bursty star formation}
\label{sec:BSF}

\begin{figure}[htbp]
\begin{center}
\subfigure{\includegraphics[width=0.46\textwidth]{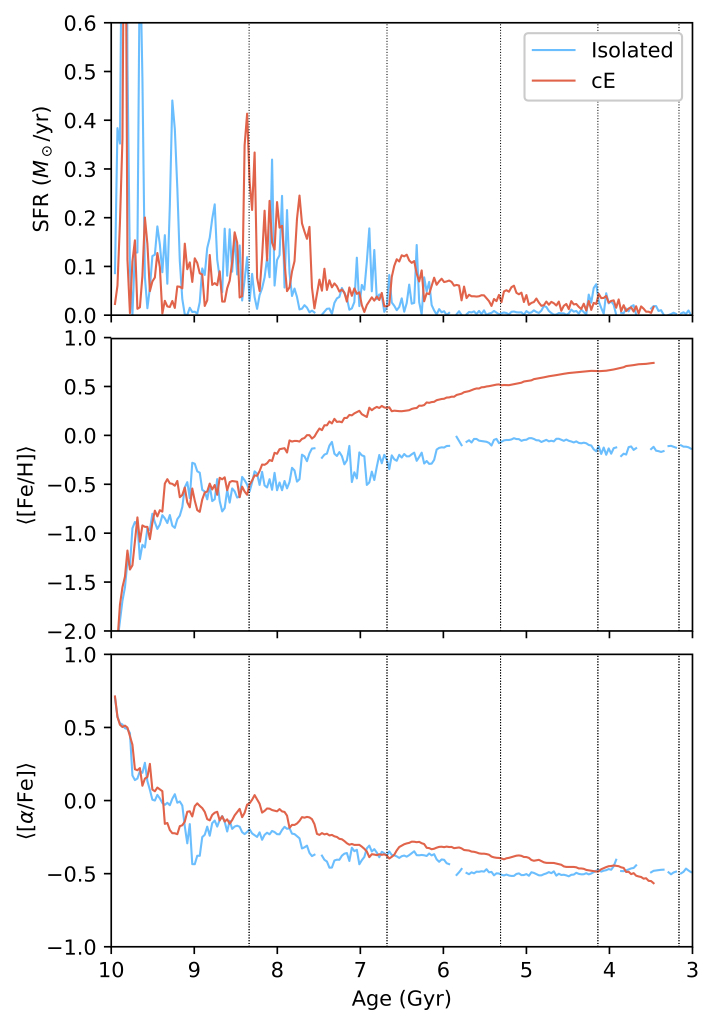}}
\caption{Evolution of the cE (red) and isolated models 
(blue) recovered from the final stellar population. From top to
bottom, the panels show the SFR, the mass-weighted mean metallicity,
$\langle$[Fe/H]$\rangle$, and $\alpha$-abundance,
$\langle$[$\alpha$/Fe]$\rangle$, as a function of age. }
\label{fig:SFR}
\end{center}
\end{figure}

\begin{figure*}[htbp]
\begin{center}
\subfigure{\includegraphics[width=0.33\textwidth]{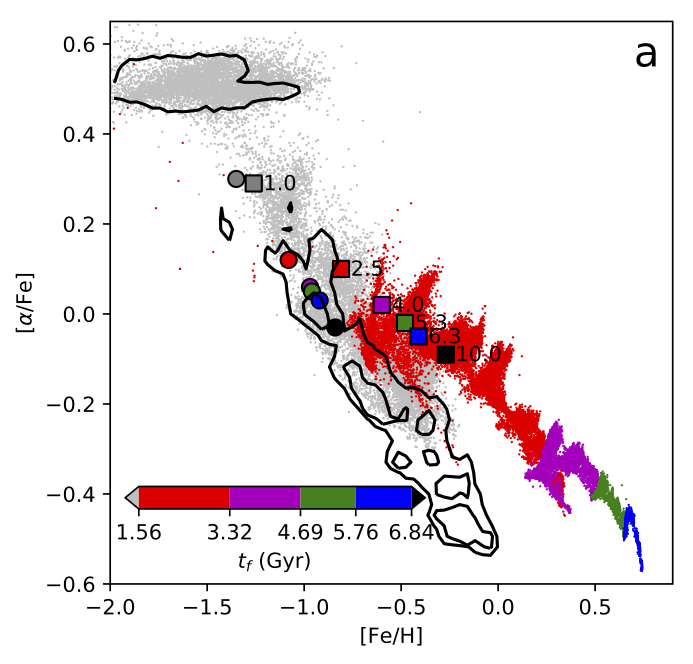}}
\subfigure{\includegraphics[width=0.66\textwidth]{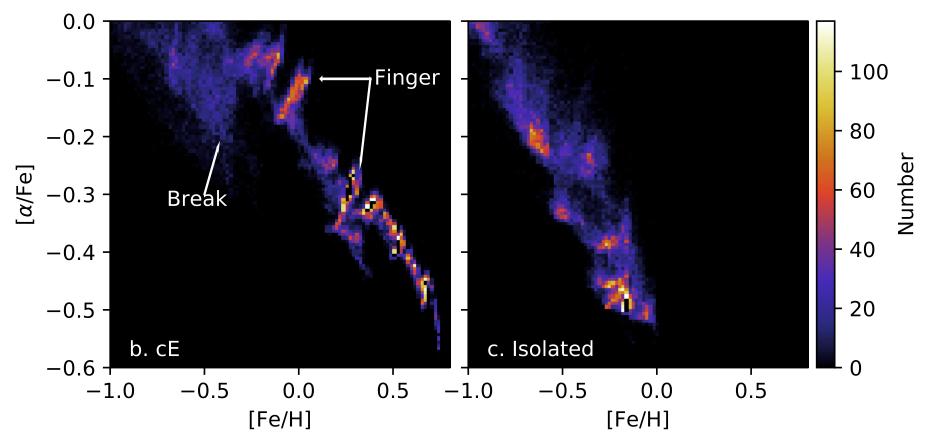}}
\caption{Stellar populations of the two models. {\bf a,} [$\alpha$/Fe] 
versus [Fe/H] for the cE and isolated models. The dots show stellar
particles from the cE at $t=10$ Gyr. These particles are binned by
their formation time, $t_\mathrm{f}$, as shown by the colorbar. The
boundaries of the colorbar are set to the times of pericentric
passages at $t=1.56, 3.32, 4.69, 5.76$, and 6.84 Gyr. Thus, each color
covers one bursty SF event. Notice that the gray dots are stars 
forming earlier than 1.56 Gyr. For comparison, the black contours show
the isolated model. The squares and circles indicate the mean
chemistry of the cE and isolated models, respectively, at the times
indicated. The chemical tracks of the cE (b) and isolated (c) models
show large differences, with a prominent break at first pericentre and
several fingers produced by bursts in the cE and a less discontinuous
evolution in the isolated model.}
\label{fig:metal}
\end{center}
\end{figure*}

In the cE model, the SFR is enhanced by compression of the ISM via
tidal interactions and high ram pressure. The top panel of
\reffig{fig:SFR} shows the SFR recovered from the stars left in the cE
at $t=10$ Gyr. The star formation history (SFH) of the isolated model
is also shown for comparison. Due to the fact that the old stars in
the cE are relatively weakly bound, they are more easily tidally
stripped compared to the young stars; indeed $65\%$ of stars formed
before $t = 1.3$ Gyr are stripped by 10 Gyr. This reduces the apparent
SFR of stars older than $8.7$ Gyr. The cE has a ``bursty'' SFH with
multiple long-lasting starbursts separated by quiescent periods, which
are consistent with its orbital period around the host. The bursty SF
events are triggered at pericentric passages (vertical dotted lines),
and last several hundreds of Myr, due to the combined effect of high
ram pressure and tidally induced compression. Such a bursty SFH was
also found in the satellite galaxies of the \Illustris\ simulation
\citep[see Fig. 7 of][]{Mistani2016}, but not in field galaxies. SF
occurs randomly in the isolated model, and rarely lasts a long time
due to the modulating influence of feedback \citep{Stinson2006}.


\section{Ram-pressure confinement}
\label{sec:gas}

Generally speaking, cEs are significantly more metal-rich than
galaxies of comparable masses. The origin of this high metallicity in
cEs is crucial for understanding their formation. Metal-rich material,
synthesized in stars, is returned via supernova feedback into the
surrounding ISM. Some of this gas is retained, enriching the star
forming gas reservoir after cooling down. New stars then reflect the
increased metallicity of the cool gas reservoir. The amount of
outflowing gas that is lost depends on the ability of the galaxy to
confine the hot gas created by supernova explosions. The shallow
gravitational potential well of lower-mass galaxies results in metals
being expelled more efficiently, producing the mass-metallicity
relation \citep{Carton2015}. \citet{Janz2016} showed that cEs fall
well above the relation between the escape velocity and metallicity
for normal galaxies \citep{Scott2009, Scott2013}. Therefore, there
must be an additional mechanism that allows cEs to retain their
supernova-enriched hot gas. In this section, we investigate the
confinement of metals in the cE model and compare it with that of the
isolated model.

\subsection{Rapid chemical enrichment via ram-pressure confinement}

The middle and bottom panels of \reffig{fig:SFR} show the
mass-weighted mean metallicity, $\langle$[Fe/H]$\rangle$, and
$\alpha$-abundance, $\langle$[$\alpha$/Fe]$\rangle$, of the stars at
$t=10$ Gyr as a function of their age. Initially, 
$\langle$[Fe/H]$\rangle$ saturates at about $-0.5$ in both the isolated
and satellite models, but after the first pericentric passage, the
subsequent star formation produces more metal-rich stars in the
satellite, while the metallicity of stars in the isolated model rises
only slightly. Just before quenching, the cE produces stars that
are $\sim 1$ dex more metal-rich than in the isolated galaxy.

\reffig{fig:metal} compares the chemical evolution of
the two models in the [Fe/H]-[$\alpha$/Fe] space. The colors in panel
({\bf a}) represent the formation times, $t_\mathrm{f}$, broadly
binned by the times of pericentric passages at $1.56, 3.32, 4.69,
5.76$, and 6.84 Gyr. Thus, each color corresponds to one bursty SF
event triggered by a pericentric passage. The small gray dots
correspond to stars with $t_\mathrm{f}<1.56$ Gyr. The black contours
show the distribution of the stars in the isolated galaxy at $t=10$
Gyr. The oldest stars ($t_\mathrm{f}<1.56$ Gyr) in the cE (gray dots
in panel (a) of \reffig{fig:metal}) follow a similar trend to those of
the isolated galaxy (black contours), while the stars forming later
follow a shallower trend (see also panels (b) and (c)). Both
Fig. \ref{fig:SFR} and \ref{fig:metal} show that stars in the cE are
noticeably more metal-rich than in the isolated model. Moreover, this
is not just true on average, since the cE is able to form more
metal-rich stars than in the isolated model.

Fig. \ref{fig:metal} also shows some unique features in the chemical
evolution of the cE galaxy. Most prominently, the evolutionary track
develops a break at first pericentre, with the star formation burst
leading to [$\alpha$/Fe] rising rapidly. Subsequent star formation
occurs on an offset track. Several bursts of star formation following
the first pericentric passage lead to prominent ``fingers,'' which
though do not alter the overall evolutionary trend until the next
pericentric passage. At later pericentric passes, star formation again
leads to rapid rises in [$\alpha$/Fe], leading to further but weaker
breaks.

\begin{figure*}[htbp]
\begin{center}
\subfigure{\includegraphics[width=0.48\textwidth]{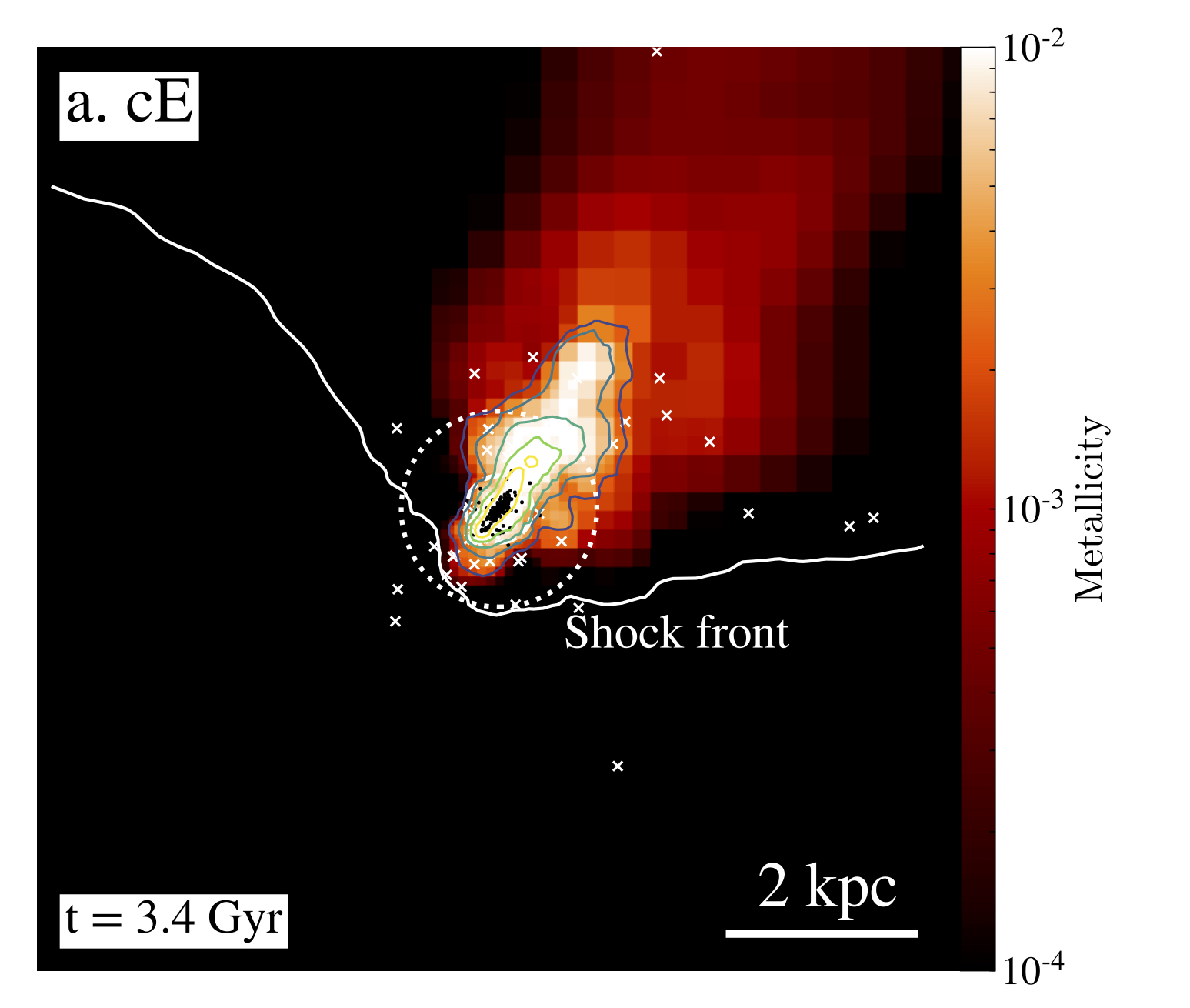}}
\subfigure{\includegraphics[width=0.48\textwidth]{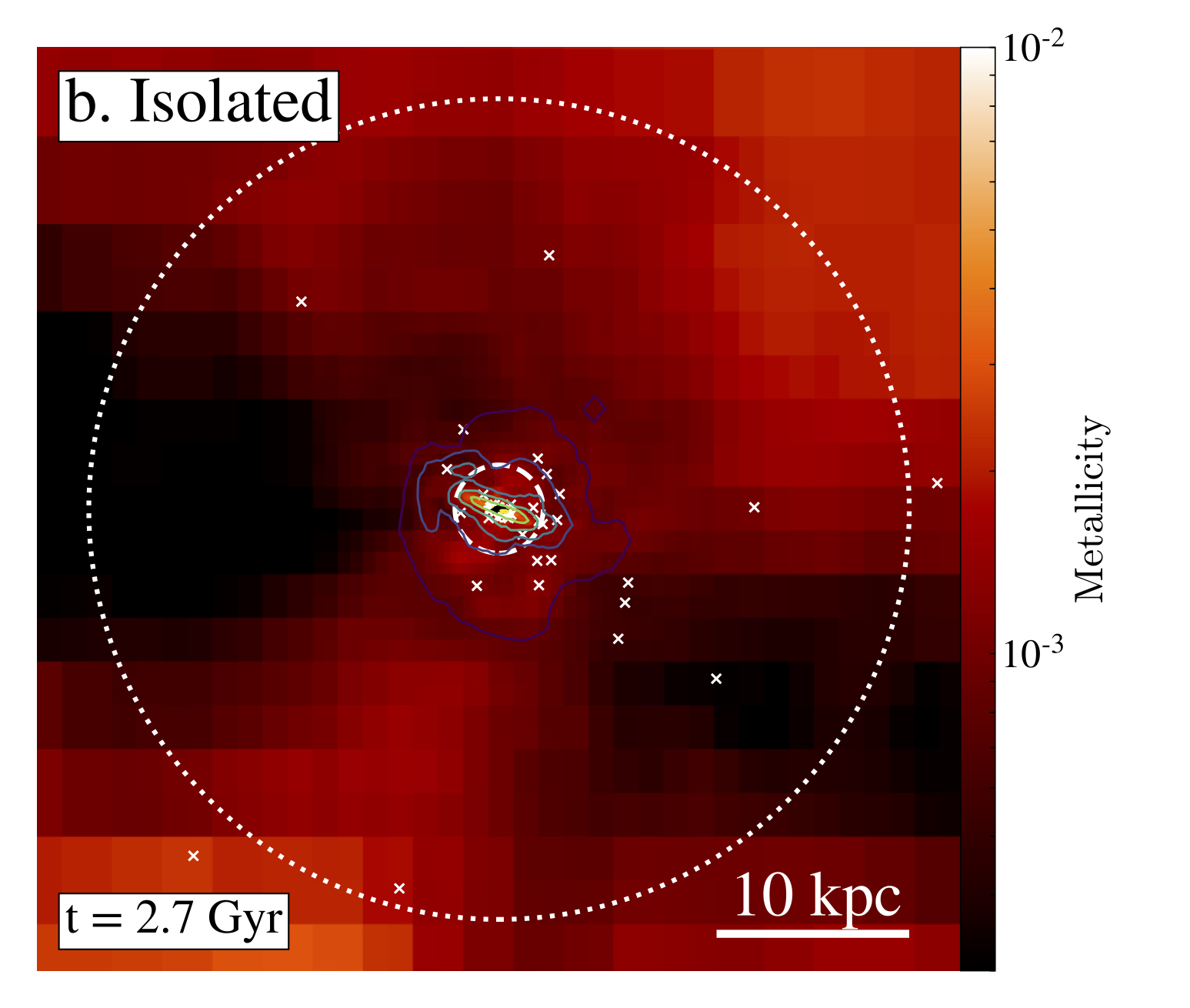}}
\caption{Fate of gas heated by supernova feedback in the two models.
Black dots within a sphere of radius $r<0.5$ kpc represent gas
particles heated to $2 \times 10^4 < T < 10^6$ K. These are selected
at $t=3.4$ Gyr in the cE ({\bf a}) and at $t=2.7$ Gyr in the isolated
model ({\bf b}). These particles are used as tracers of metal-rich
outflows produced by supernovae. The white crosses mark their
positions 0.5 Gyr later. The dashed and dotted circles represent
$r_\mathrm{50}$ and $r_\mathrm{80}$, the radii containing $50\%$ and $80\%$
of the chosen particles 0.5 Gyr later. Gas density contours
are shown in blue and the background indicates the fractional metal
abundance of the gas. The thick white contour in panel {\bf a}
corresponds to the ram-pressure shock front. Note the different scale
of the two panels.}
\label{fig:gasmotion}
\end{center}
\end{figure*}

\begin{figure*}[htbp]
\centering
\includegraphics[width=0.48\textwidth]{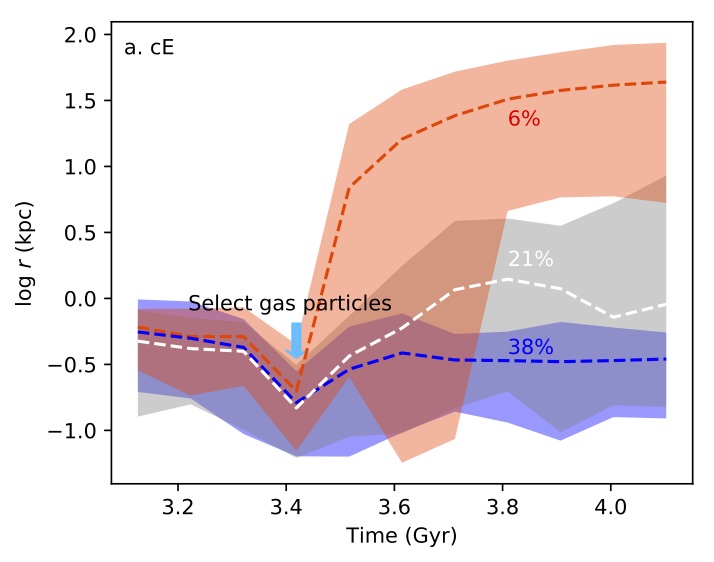}
\includegraphics[width=0.48\textwidth]{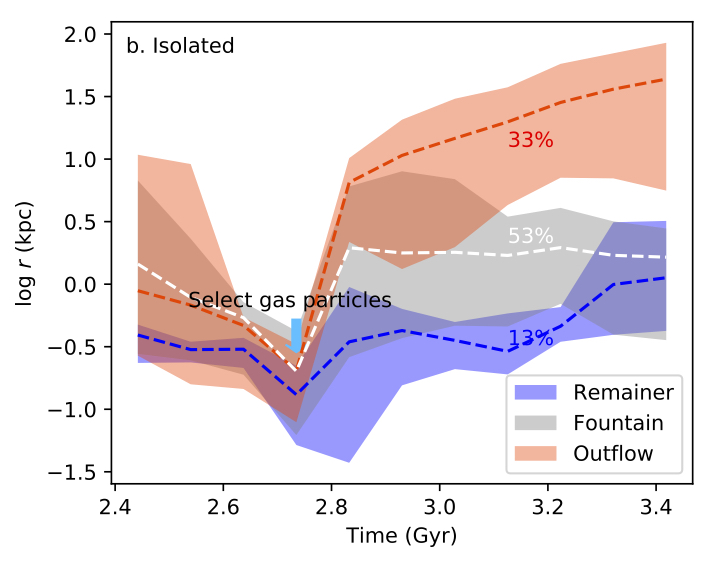}
\caption{Evolution of the radial distribution of the hot
gas particles heated by supernovae. We classify the particles into
three groups according to their locations 0.5 Gyr later. The gas
particles escaping to $r>5$ kpc are classified as ``outflow'' (red),
while the members of the ``remainer'' group (blue) are confined within
$r<1$ kpc. Except for particles depleted by SF, the rest of the gas
particles belong to the ``fountain'' group. The dashed lines represent
the mass-weighted mean radius in each group; the shaded regions
correspond to the $1\sigma$ envelopes. The mass fractions of each
group are given in terms of percentage. About $35\%$ and $1\%$ of the
selected gas particles are depleted by SF in the cE and isolated
models, respectively.}
\label{fig:Gasradial}
\end{figure*}

The environment around the massive galaxy therefore plays an important
role in not only triggering SF but also retaining the metals produced
in supernovae. In Figs. \ref{fig:gasmotion} and \ref{fig:Gasradial},
we trace the gas particles heated by supernova explosions to follow
the metal-rich outflows. We consider the gas particles that are heated
to $>20, 000$ K from $<15, 000$ K within a short time ($<10$ Myr here) when 
starbursts are on-going. Only the hot gas particles located within 
0.5 kpc of the center of the dwarf galaxy are used. Such gas particles 
are generally more metal-rich than their surrounding environment 
as they obtain not only energy but also metals from 
supernovae. In the isolated model, we identify 71 such gas particles, for a total 
mass $\simeq 1.6 \times 10^6 M_\odot$. The longer lasting starburst in the
cE heats more gas to high temperatures. As a consequence, 235 heated gas 
particles (mass $\simeq 5.2 \times 10^6 M_\odot$) are identified.

In \reffig{fig:gasmotion}, we mark the projected position of such gas
particles (black dots). The white crosses represent the positions of
the heated gas particles 0.5 Gyr later. We define the radius
containing $50\%$ of these particles 0.5 Gyr later as
$r_\mathrm{50}$, corresponding to the dashed circles in
\reffig{fig:gasmotion}. Similarly, $r_\mathrm{80}$ (dotted circles)
represents the radius containing $80\%$ of the heated
particles. $r_\mathrm{50}$ and $r_\mathrm{80}$ of the selected gas
particles in the cE are 0.4 and 1.5 kpc, respectively, which are substantially
smaller than in the isolated model (2.6 and 22.2 kpc). As a result,
the metal-rich gas is less efficiently expelled by feedback in the cE
model.
In the cE, the outskirts of the gas
distribution is stripped backward by ram pressure
(\reffig{fig:gasmotion} (a)). In comparison, metal-rich outflows are
present perpendicular to the disk (\reffig{fig:gasmotion} (b)) in the
isolated model.

\begin{figure}[htbp]
\begin{center}
\subfigure{\includegraphics[width=0.48\textwidth]{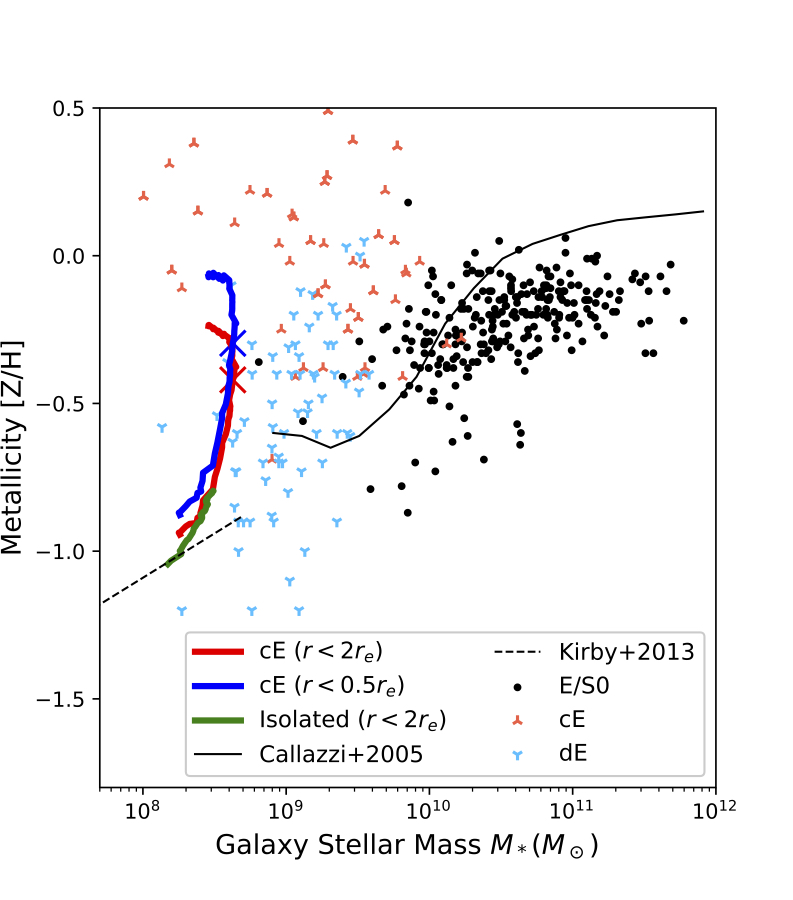}}
\caption{Evolution of the cE and isolated models overlaid on the 
observed mass-metallicity relation. The data of E/S0s (black dots),
cEs (red stars) and dEs (blue stars) are adopted from
\citet{Janz2016}. The black solid and dashed lines correspond to the
mass-metallicity relations of \citet{Gallazzi2005} and
\citet{Kirby2013}, respectively. The metallicity is measured in more
central regions in \citet{Gallazzi2005}, which leads to a higher
metallicity than those of \citet{Janz2016}. The red and blue tracks
represent the chemical evolution of the cE model measured from stars
within the 2 and 0.5 effective radii, respectively. The crosses mark $t=4$
Gyr, which separates the two stages of the evolution of the cE. The
green track is the chemical evolution of the isolated model.}
\label{fig:ZHob}
\end{center}
\end{figure} 
 
\reffig{fig:Gasradial} follows the evolution of the radial distribution 
of the selected gas particles. We classify them into ``outflow''
(red), ``fountain'' (gray), and ``remainer'' (blue) groups. The gas of
the outflow group escapes to $r>5$ kpc, while the members of the
remainer group never move beyond 1 kpc from the center of the
dwarf. The rest of the gas particles, except for depleted ones, belong
to the fountain group. Their mean radii are shown by dashed lines, and
the shaded regions correspond to their $1\sigma$ envelopes. In the cE,
a majority of the gas particles are confined to the central region,
where they can form new stars. About $35\%$ of the selected gas
particles are completely depleted by SF during this
period. Furthermore, many fountain gas particles fall back to the
center. Only $\sim 6\%$ of the selected gas particles are lost, in the
direction opposite to the cE's bulk motion, where the ram pressure is
low. The majority of the gas is confined to a relatively small volume
by the ram pressure, allowing rapid enrichment in the cE. In contrast,
most of the hot gas particles in the isolated model escape to $>1$
kpc. As a consequence, the stellar metallicity of the isolated model
increases much more slowly.

\reffig{fig:ZHob} compares the stellar mass and metallicity of the
models to observations. The cE's metallicity increases rapidly during
stage (1) ($t<4.0$ Gyr). Considering that the measured metallicity is
affected by the region observed, we show the metallicity within 0.5
and 2 effective radii. The metallicity measured at the central region
is higher because of the steep radial gradient. The galaxy evolved in
isolation is significantly less metal rich.

It is clear that the high ram pressure plays a crucial role in
confining the metals released in supernova events. {\it Under high ram
pressure, metals are synthesized and recycled highly efficiently. This
allows new stars to reach a high metallicity by confining the
metal-rich gas in close proximity to the cE.}

\subsection{The effect of tidal stripping}

Due to the fact that the old stars (those formed before the first
pericentric passage) in the cE are relatively weakly bound, they are
easily stripped with respect to the young stars forming in the central
region. Since old stars are more metal-poor and $\alpha$-enhanced,
this loss enhances the average metallicity further, while reducing
[$\alpha$/Fe]. $65\%$ of stars forming before $t=1.3$ Gyr are stripped
by 10 Gyr. As a consequence, the overall metallicity of the cE keeps
increasing during stage (2) from 4 Gyr to 10 Gyr, as seen in
Fig. \ref{fig:ZHob}. In comparison, the metallicity in the isolated
model changes little after 4 Gyr. Tidal stripping reduces the
$\alpha$-abundance faster in the cE than in the isolated model
(\reffig{fig:metal}), although the stars forming via the bursty SF in
the cE are more $\alpha$-abundant (see the third panel of
\reffig{fig:SFR}).


\section{Discussion}
\label{sec:dis}

The scenario presented here relies on cEs forming in close proximity
to a larger halo. This scenario is likely to be even more efficient in
cluster environments where a low-mass satellite can reach a higher
velocity, and the confining corona is hotter. 
The existence of cEs in the field \citep{Huxor2013, Paudel2014a,
Chilingarian2015} is therefore likely a consequence of ejection in a
three-body encounter between the cE, the host galaxy, and a third
galaxy \citep{Chilingarian2015}. 

The broad metallicity distribution of low-mass galaxies suggests that
the effect of environment is important. cEs are significantly offset
from the normal mass-metallicity relation. We suggest that high ram
pressure in high-speed passages around massive host galaxies confines
metals in low-mass galaxies just when star formation is enhanced
because the gas is also compressed. To test the effect on the metallicity
of the subgrid feedback; we tested the early part of the evolution
using a stronger feedback model \citep[super bubble
feedback,][]{Keller2014} and found that even in that case the
supernova ejecta are confined by ram pressure during the closest
passage. Independently, \citet{Williamson2018} used wind tunnel
experiments of dwarf galaxies to also demonstrate that ram pressure
confines gas, and metals, to dwarf galaxies.

\subsection{Comparison with previous results}

Many studies have addressed the formation and evolution
of low-mass dwarf satellites in the vicinity of a massive host
\citep[see the review][]{Mayer2010} and references therein). Ram pressure is
often shown to strip gas dramatically in dwarf
galaxies, then their disks are tidally heated to dSphs. In
a cosmological context, \citet{Sawala2012} simulated the dwarf
satellites in Milky-Way sized halos, finding a full family of
dSphs/dEs formed. However, none of these studies have ever reported
cEs produced by gas-rich dwarfs falling into the vicinity of a massive
host. The absence of cEs is possibly because previous simulations have
not, in general, used very eccentric orbits. Moreover, the resolution 
of cosmological simulations is possibly unsufficient in resolving cEs.

An exception was the work of \citet{Kazantzidis2017}, who studied the
evolution of disky dwarfs orbiting Milky Way-sized hosts. They used
pre-existing disks of varying gas fractions to represent the initial
dwarfs; these had initial scale length of $\thickapprox 0.76$ kpc,
making them already extended dwarfs. In agreement with our results,
they find that the initially gas-rich dwarf on a highly plunging 
orbit (their model S14) ends up slowly rotating, spherical, and gas-free.
While they identify such galaxies as dSphs, \citet{Kazantzidis2017} do not
discuss the compactness and metallicity of their dwarfs, and therefore it unclear
whether they also form cEs under ram pressure confinement. However, an
important difference in their models is that they already start with a
quite extended disk. 

In this paper, we present a possible scenario for the formation of metal-rich cEs from gas-rich, 
low-mass galaxies. In this scenario, a highly radial orbit of the satellite plays an important 
role in triggering a starburst and confining metals in the vicinity of a massive host. It is unclear 
what roles the mass and spin parameter $\lambda$ of the satellite play. Future work will 
explore a broader range of initial conditions. However, the differential comparison we have 
performed between the isolated dwarf on the one hand and the satellite dwarf on the other, 
both of which have exactly identical initial conditions, suffices to highlight the physical influence 
of the environment. 

Using a more massive and slowly spinning initial satellite will 
possibly favor the formation of cEs as well. In this work, we use $\lambda = 0.02$.
This is lower than the median value of $\sim 0.04$ \citep[e.g.][]{Bett2007}, although 
$\lambda$ has a very large scatter in low-mass halos, varying from 0.01 to 1. 
Moreover, \citet{Rodriguez-Puebla2016} argue that $\lambda$ at high redshifts is lower by a 
factor of 2 than that at low redshifts. Thus, our choice of $\lambda$ is reasonable 
if the cE falls in at high redshifts. A systematic study is still required.

\subsection{Observational consequences}

As shown in \reffig{fig:ZHob}, the cE model at 10 Gyr exhibits
comparable metallicity to observed cEs, while the isolated model is
consistent with observed dEs. Furthermore, many cEs exhibit moderate,
or even sub-solar, [$\alpha$/Fe] (the bottom panel of Fig. 5 in
\citet{Janz2016}. In this scenario, it is reasonable that cEs exhibit
very high metallicities but a wide range of [$\alpha$/Fe] due to
the combined effect of ram-pressure confinement, tidal stripping, and
bursty SF.

The trend of [$\alpha$/Fe] and [Fe/H] can be used to constrain the
formation of cEs. The break corresponds to the time that the cE first
passes pericentre, while the ``fingers'' represent bursts of star
formation after the pericentre. In between bursts, because the dwarf's
gas is confined to a small volume, it is chemically well mixed, and
stars form with a relatively small scatter. These features, being of the
order of $\sim 0.1$ dex in $\alpha$-abundance, can be tested with future
observations of M32.


\section{Summary}
\label{sec:sum}
We used a high-resolution simulation to investigate the evolution of a
low-mass $\sim 10^8 M_\odot$ galaxy infalling on a highly radial orbit
around a massive disk galaxy. A compact, metal-rich galaxy forms
quickly due to the close flybys. We propose that a gas-rich, diffuse
normal low-mass galaxy is replaced by a metal-rich cE within several 
Gyr. The evolution of the galaxy's structure and
metallicity can be separated into two stages:

\begin{itemize}
\item[(1)] While a large fraction of gas is stripped by ram pressure, 
the remaining gas in the centre sustains bursty star formation
triggered by the combined effects of ram pressure confinement and
tidal compression at pericentric passages. During periods of high ram
pressure, the metal-rich outflows driven by supernovae and stellar
winds are significantly suppressed. Thus, the infalling galaxy is able
to retain more metals than in isolation, leading to the formation of a
cE galaxy with high [Fe/H].

\item[(2)] Following the quenching, frequent strong tidal interactions 
slowly change the mass, morphology, and metallicity of the cE. The
diffuse outskirts of the cE, composed mainly of old metal-poor and
$\alpha$-rich stars, are tidally stripped. The cE becomes less
massive, more metal-rich, but less $\alpha$-enhanced due to tidal
stripping. Tidal stirring gradually transforms the compact galaxy into
a nearly isotropic object that is dominated by random motions.
\end{itemize}

The fact that cEs fall above the well-known mass-metallicity relation
\citep[e.g.][]{Janz2016} has been considered as strong evidence that
cEs are stripped remnants of massive galaxies. Indeed, cEs embedded in
tidal streams have been considered as ``smoking guns'' of the tidal
stripping scenario \citep{Huxor2011}. However, we have shown here that
although tidal stripping plays a role in the formation of cE galaxies,
it is not the most important mechanism. By tracing gas particles
heated by supernovae, we verified that a high ram pressure environment
confines more metals than when evolved in isolation. We suggest
that the rapid metallicity enrichment in cEs is a natural outcome of
the suppression of outflows and bursty SF for satellite galaxies
orbiting in a dense corona around a massive host. The ram pressure
confinement described here results in a very distinct and testable
evolutionary track in [$\alpha$/Fe]-[Fe/H] space, with a shallower
gradient following the cE's first pericentric passage; ``finger-like''
projections corresponding to SF bursts; and abrupt offsets to higher
[$\alpha$/Fe] and [Fe/H] at pericenters. This scenario implies that the 
progenitor does not need to be much
more massive than the cE itself; as a result, the debris is likely to
amount to a comparable mass as the satellite itself. 

Ram-pressure confinement of metals likely affects galaxies in various 
types of dense environment, such as the vicinity of massive galaxies, rich groups, 
or clusters.  Ram pressure stripping has received appreciable attention in the past. 
It is an efficient mechanism for stripping gas and suppressing SF, especially in the 
outer regions of satellite galaxies where gas is less bounded. Our results suggest that 
high ram pressure in the central regions of satellites sustains SF and leads to the build 
up of a metal-rich core. This effect might be more important in low-mass satellites 
whose gravitational potential is otherwise not deep enough to maintain metals.
It sheds new light on the evolution of satellite galaxies.

\begin{acknowledgements}
This work was supported by the National Key R\&D Program of China
(2016YFA0400702) and the National Science Foundation of China
(11473002, 11721303). The simulations in this paper were run at the
DiRAC Shared Memory Processing system at the University of Cambridge,
operated by the COSMOS Project at the Department of Applied
Mathematics and Theoretical Physics on behalf of the STFC DiRAC HPC
Facility (www.dirac.ac.uk). This equipment was funded by BIS National
E-infrastructure capital grant ST/J005673/1, STFC capital grant
ST/H008586/1 and STFC DiRAC Operations grant ST/K00333X/1. DiRAC is
part of the National E-Infrastructure. The analysis was performed
using yt \citep[http://yt-project.org,][]{Turk2011}. This project is
also supported by the High-performance Computing Platform of Peking
University. M.D. is supported by the grants ``National Postdoctoral
Program for Innovative Talents'' (\#8201400810) and ``Postdoctoral 
Science Foundation of China'' (\#8201400927) from the China
Postdoctoral Science Foundation. V.P.D. was supported by STFC
Consolidated grant ST/R000786/1. We also acknowledge support 
from a Newton Advanced Fellowship no. NA150272 awarded by 
the Royal Society and the Newton Fund.

Finally, I (M.D.) warmly thank my wife Pan Zhang for constant support and 
love, even when she was pregnant. Fan-Yu Du, my son, hope you enjoy your 
journey to the universe, just like your name.

\end{acknowledgements}

\bibliographystyle{apj}
\bibliography{cE_cit}

\end{document}